\documentstyle[preprint,aps]{revtex}

\begin{document}
\title{Structures and melting in infinite gold nanowires}
\draft
\tightenlines
\author{G. Bilalbegovi\'c}
\address{Department of Physics, University of Rijeka, 
Omladinska 14, 51000 Rijeka, Croatia}
\date{to be published in Solid State Communications}
\maketitle

\begin{abstract}
The temperature dependence of structural properties for infinitely long gold
nanowires is studied. The molecular dynamics simulation method and the
embedded-atom potential are used. The wires constructed at $T=0$ K with a
face-centered cubic structure and oriented along the (111), (110), and (100)
directions are investigated. It was found that multiwalled structures form
in all these nanowires. The coaxial cylindrical shells are the most
pronounced and well-formed for an initial fcc(111) orientation. The shells
stabilize with increasing temperature above $300$ K. All nanowires melt at
$T<1100$ K, i.e., well below the bulk melting temperature.

\end{abstract}


Metallic nanowires are important for applications and for understanding of
fundamental properties of materials at nanoscales. Over the past several
years investigations on metallic nanowires were devoted mainly to the
properties of cylindrical junctions formed between a metallic tip and a
metallic substrate \cite{Landman}. Gold nanostructures whose diameter and
length are about $1$ $nm$ were recently formed in a scanning tunneling
microscope and studied by a high-resolution electron microscope \cite
{Ohnishi,Leiden}. Unusual vertical rows of gold atoms were observed.
Therefore, it is important to study the internal structure of gold
nanowires. Computer simulations are suitable for these investigations.

An important topic in the cluster science is the melting of nanoparticles 
\cite{Pawlow,Borel,Martin}. 
Experimental, theoretical, and computer simulation studies
have shown that the melting temperature depends on the cluster size. These
studies suggest the dependence of the form: 
\begin{equation}
T_m = T_b - c/R,  \label{eq:1}
\end{equation}
where $T_m$ is the melting temperature for the spherical nanoparticle of
radius $R$, $T_b$ is the bulk melting temperature, and $c$ is a constant.
In recent studies deviations from this law for small sizes
are found \cite{Borel,Martin}.
Melting of clusters, i.e., spherical nanoparticles, was the subject of
several recent experiments \cite{Martin}. In contrast, melting of nanowires
was not studied experimentally. The exception is an early work on mercury
filaments with diameters between $2$ and $10$ $nm$ \cite{Bogomolov}. In this
experiment a decrease in the melting temperature was detected from the
resistance measurements.

It is well known that the presence of geometric (i.e., atomic) and
electronic shells determines various properties of clusters \cite{Martin}.
Electronic shells in finite sodium nanowires 
were recently found in a jellium model
calculation \cite{Yannouleas}.
These shells  were also observed in the conductance
measurements \cite{Yanson}. 
The Molecular Dynamics (MD) simulation has shown an existence 
of multishelled finite gold nanowires at room
temperature \cite{Goranka}. 
The cylindrical shells obtained in this simulation 
resemble geometric shells in
clusters. Infinite wires with periodic boundary conditions along the wire
axes are more often studied by MD simulations. 
For example, structures of ultra-thin infinite Pb
and Al nanowires at $T=0$ K were studied by MD simulation \cite{Oguz}.
In comparison with finite nanowires,
infinite wires are stable in simulations at high temperatures. 
The MD method was
used to investigate premelting of infinite Pb nanowires with the axes along
a (110) direction \cite{Gulseren}. The size-dependence of melting properties
for cylindrical and spherical geometry was compared. It was found that the 
$1/R$ behavior, as in Eq. (1), 
approximately holds for a nanowire of radius $R$. 
Melting of platinum and silver infinite (100) oriented nanowires was also
studied by the MD method \cite{Finbow}. A decrease in the melting
temperature was investigated, as well as a change in energy on melting. In
Ref. \cite{Goranka} finite gold nanowires with length of up to several $nm$
and axes initially oriented only along the (111) direction were investigated
at $T=300$ K. It is also interesting to study infinite gold nanowires. 
These studies should explain 
whether multishelled structures appear in infinite wires and starting from
other orientations of the axes. Simulations at various temperatures may
suggest the optimal method for the fabrication of nanowires. Studies of
melting in multiwalled nanowires may also shed light on the role of
shells in melting of clusters.
In this paper a MD study of the structural and melting properties for
infinite gold nanowires is presented. These infinite wires, in comparison
with finite gold nanowires in Ref. \cite{Goranka}, resemble longer
nanowires fabricated in the laboratories. For example, the electron-beam
litography methods nowadays produce metallic nanowires whose length is 
$1\mu$m and more.

Nanowires were simulated using the classical MD method. The embedded-atom
potential with proven reliability for modeling properties of gold was
employed \cite{Furio}. Several fcc (111), (100), and (110) oriented gold
nanowires were prepared at $T=0$ K and then a procedure for a wire
preparation from Ref. \cite{Gulseren} was used. 
Circular cross-sections with maximal
radii of $0.9$ and $1.2$ $nm$ were constructed.
The periodic boundary conditions were applied along the axis direction.
The number of atoms in MD boxes was between 
$566$ and $1116$, as shown in Table I. 
The temperature was controlled by rescaling particle velocities. 
A time step of $7.14 \times 10^{-15}$
{\it s} was used. Structures were analyzed after long equilibration of $10^5-10^6$
time steps. Figures 1-5 show the results obtained after evolution of $10^5$
time steps.

Nanowires with initial fcc(111) cross-sections are labeled as $A$ and $B$ in
Table I. 
Already at low temperatures they form multiwalled structures.
A nanowire $A$
consists of three cylindrical walls and a thin core. 
The infinite nanowire $A$ at 
$300$ K has a similar structure as a finite nanowire with the same
diameter simulated at this temperature and described in Ref. \cite{Goranka}.
The same result, i.e., that only minor differences exist between finite and
infinite wires, was obtained in the first principles density functional
studies of monoatomic gold nanowires \cite{Portal,DeMaria}.
At higher temperatures 
($T<900$ K) the walls are more homogeneous than at $T=300$ K. The structure
of the nanowire $A$ at $T=800$ K is shown in Fig. 1(a).  A nanowire $B$
at $300$ K consists of a large fcc(111) core and an outer cylindrical wall.
Above $600$ K this core breaks into several shells. Figure 1(b)
shows that the nanowire $B$ at higher temperatures consists of four
cylindrical walls and a thin filled core.

Nanowires with initial fcc(110) cross-sections are labeled as $C$ and $D$ in
Table I. The nanowire $C$ consists of three shells shown in Fig. 2(a).
The core of the nanowire $C$ is square and empty. The multiwalled structure
is less pronounced in the nanowire $D$. Figure 2(b) shows that for this
nanowire the fcc structure of a core remains, although outer coaxial walls
are also formed.

Simulation for initial cross-sections
with the density and structure of unreconstructed Au(100) surface
have shown that a strong torsion of nanowires exists. Then nanowires with
the density increase in the top and bottom layers of the MD box 
were prepared. The density increment in
these layers was taken in accordance with the properties of reconstruction
on Au(100) \cite{Ercolessi,Erio}. Although the MD box repeats along 
the wire axis by the periodic boundary conditions, it was found that 
this density increase is sufficient to stabilize the nanowire.
These nanowires with a stable untwisted
cylindrical morphology are labeled as $E$ and $F$ in Table I. The nanowire 
$E$ shown in Fig. 3(a) consists of three shells. The most interior of these
shells is empty and square. It is interesting to point out a similarity
between nanowires $C$ (Fig. 2(a)) and $E$ (Fig. 3(a)) at $T=800$ K. 
As already explained, these
nanowires evolved from different initial structures at $T=0$ K. Figure 3(b)
shows that the nanowire $F$ at $T=800$ K consists of four shells. Its core
is empty, interior shells are irregular, and the walls are more disordered
than in other nanowires at this temperature. The cylindrical walls
in the nanowire $F$
are more ordered at $900$ K than at $800$ K. The structure of this
nanowire at $800$ K was presented in Fig. 3(b) for a comparison with other
nanowires. 

The cylindrical shells (such as these shown in Figs. 1-3) represent radial
density oscillations in nanowires. A related problem of a fluid in a finite
geometry was studied within the one-dimensional lattice gas model \cite
{Tarazona}. Near the walls the layering effects were found.
The inner part of the fluid behaves
as a bulk if distances between the walls are large. 
For small separations of the order of several interatomic
distances strong density inhomogeneities appear. These density oscillations
were obtained as a result of constructive and destructive interference
between ordering induced by the walls.
The MD boxes for nanowires consist of cylindrical gold surfaces at
small radial separations which 
produces strong density oscillations.
The layering effects are pronounced in gold, as well as in graphite and
other materials for which multishelled nanostructures were observed 
\cite{Goranka}.
The intershells spacings in Figs. 1-3 (and in other nanowires not
shown here) are inhomogeneous and change between $0.1$ and $0.2$ $nm$.

All nanowires melt in the temperature interval $(900-1100)$ K,
i.e., well below the bulk melting
temperature for this potential ($T_{b}\sim 1350$ K \cite{Furio}). 
With increasing temperature highly diffusive atoms progressively appear 
in all cylindrical
walls. Therefore, all shells melt simultaneously. 
The average
mean-square displacements for three nanowires are presented in Fig. 4. The
graphs for remaining three nanowires are similar. The displacements sharply
increase at $(900-1000)$ K. The diffusion coefficients have a similar
temperature dependence. The liquid nanowires up to the bulk melting temperature
retain a layered structure with limited interdiffusion.

Internal energy as a function of temperature for
different nanowires is shown in Fig. 5. Melting 
is the first-order phase transition and its typical feature 
in three-dimensional systems is a jump in the
caloric curve. In Fig. 5 the jumps in $E(T)$ are less pronounced than in
Ref. \cite{Gulseren} for the (110) oriented filled
lead nanowires with radii
between $1.3$ and $2.5$ $nm$. It is known that strict phase transitions do not
exist for one-dimensional systems \cite{Landau}. However, a possibility for
the first-order phase transitions was studied in the one-dimensional model 
\cite{Scalapino}. For this model 
pseudo-first-order transitions were found for the
cylindrical systems with cross sections of $10-20$ nearest neighbors units.
These pseudo-transitions are indistinguishable from real ones. 
Simulated here multishelled gold nanowires with smaller cross sections 
are close to one-dimensional systems. 
Filled lead nanowires with larger cross sections studied
in Ref. \cite{Gulseren} are in the regime of the pseudo-first-order
transitions and the jumps in $E(T)$ are apparent. 
Approximate finite-size scaling
analysis shows that a first-order phase transition at temperature $T_{c}$
in a finite system is broadened over an interval: 
\begin{equation}
\frac{\Delta T_{c}}{T_{c}}\sim \frac{1}{N\sigma },  \label{eq:2}
\end{equation}
where $\sigma$  is the latent entropy of the bulk transition and $N$ is the
number of particles \cite{Imry,Binder}. 
The latent entropy in the Eq. (2) 
is the ratio between a latent heat per particle
and $T_c$, measured in units of the Boltzmann constant.  
The latent heat of melting for Au 
is $12.54$ kJ/mol \cite{www}.
Equation (2) gives that the broadening of the transition for gold
nanowires is $\Delta T_c/T_c \sim 10^{-3}$. 
Figure 5 shows that the transition is smeared over larger temperature
intervals. Additional contribution to the broadening of the melting 
transition is given by the decrease of the latent heat with the size.
The latent heat of this transition tends to
disappear as the size of the system decreases. The melting of multishelled
nanowires is also the structure dependent. 
Therefore, because of the irregularity of the shells the features of
the melting transition 
for multishelled nanowires do not change regularly with the number of atoms.

In conclusion, a MD study of structural and melting properties for infinite
gold nanowires within the framework of the embedded-atom method is
presented. It was found that in all nanowires coaxial cylindrical walls
exist. The shells are the most pronounced for an initial fcc(111)
orientation. The walls are the most homogeneous at $(800-900)$ K.
Nanowires melt around $1000$ K by simultaneous melting of all shells.
This study suggests the optimal temperature and the initial structure for
fabrication of multishelled gold nanowires. These nanowires are important 
for applications in mesoscopic electronic devices. The melting of 
multiwalled nanowires brings an insight into a specific first-order phase
transition in unusual multiwalled quasi one-dimensional structures.

\acknowledgments
This work has been carried out under the CRO-MZT project
119206-``Dynamical Properties of Surfaces'' and
the EC Research Action COST P3-``Simulation of Physical Phenomena in
Technological Applications".

\begin{table}
\caption{ Geometry of investigated nanowires. The radius here is the maximal
distance from the wire axis in an initial fcc structure.}
\label{table1}
\begin{tabular}{llll}
Nanowire & Initial fcc cross-section & Radius ({\it nm}) & Number of atoms \\ 
\hline
$A$ & $(111)$ & $0.9$ & $588$ \\ 
$B$ & $(111)$ & $1.2$ & $1032$ \\ 
$C$ & $(110)$ & $0.9$ & $612$ \\ 
$D$ & $(110)$ & $1.2$ & $1116$ \\ 
$E$ & $(100)$ & $0.9$ & $566$ \\ 
$F$ & $(100)$ & $1.2$ & $952$
\end{tabular}
\end{table}

\begin{figure}
\caption{ Structure of nanowires $A$ (a) and $B$ (b). The trajectory plots
refer to $800$ K, a time span of $\sim 7$ {\it ps}, 
and include all atoms in the
slice of $4$ {\it nm} along the axes of the cylindrical MD box. 
The geometry of all nanowires is described in the text and Table I.}
\label{fig1}
\end{figure}

\begin{figure}
\caption{ Structure of nanowires $C$ (a) and $D$ (b). 
The trajectory plots are prepared
as in Fig. 1.}
\label{fig2}
\end{figure}

\begin{figure}
\caption{ Structure of nanowires $E$ (a) and $F$ (b). The plots are prepared
as in Fig. 1.}
\label{fig3}
\end{figure}

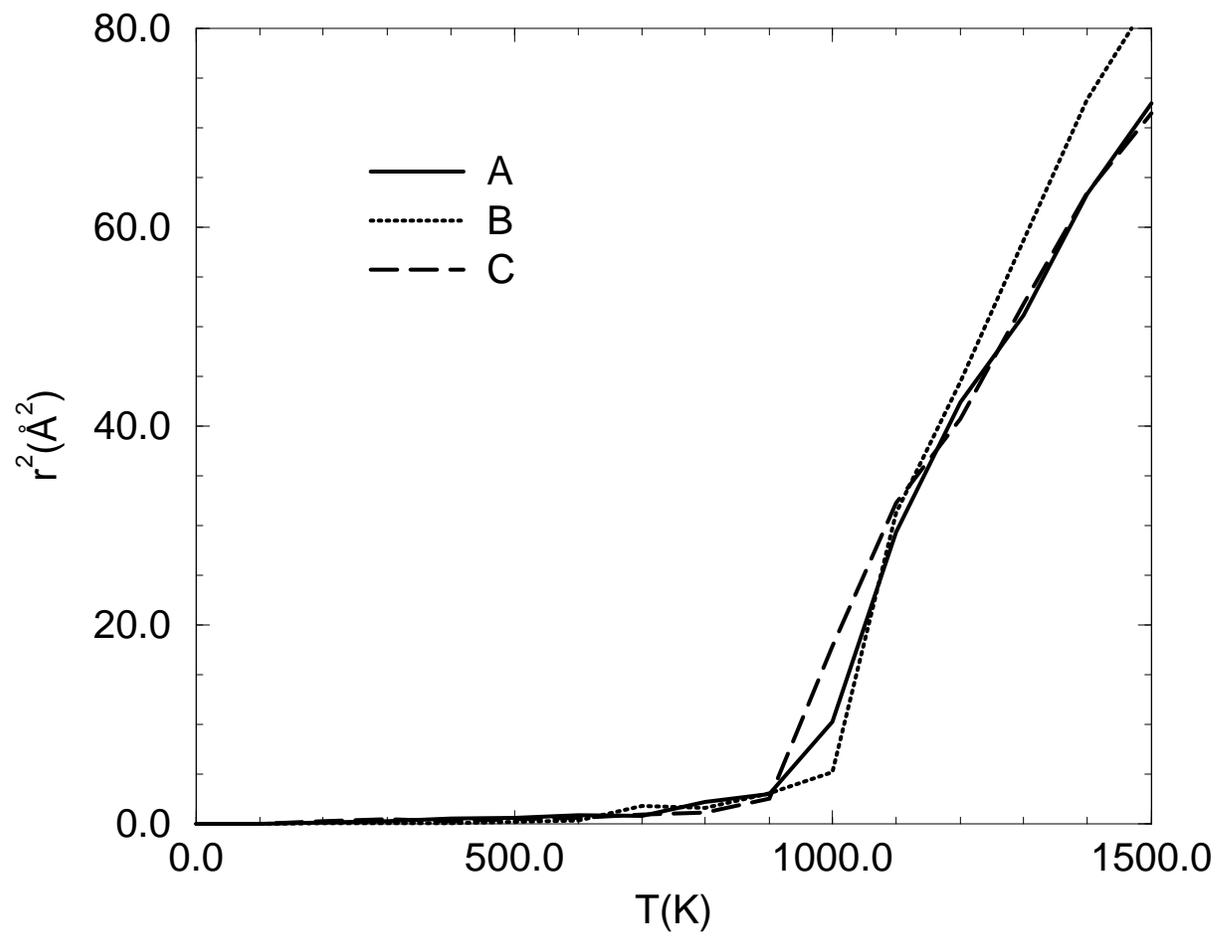
\begin{figure}
\caption{ Mean-square displacements as a function of temperature for
nanowires $A$, $B$, and $C$, after $0.714$ $ns$ of time evolution.}
\label{fig4}
\end{figure}

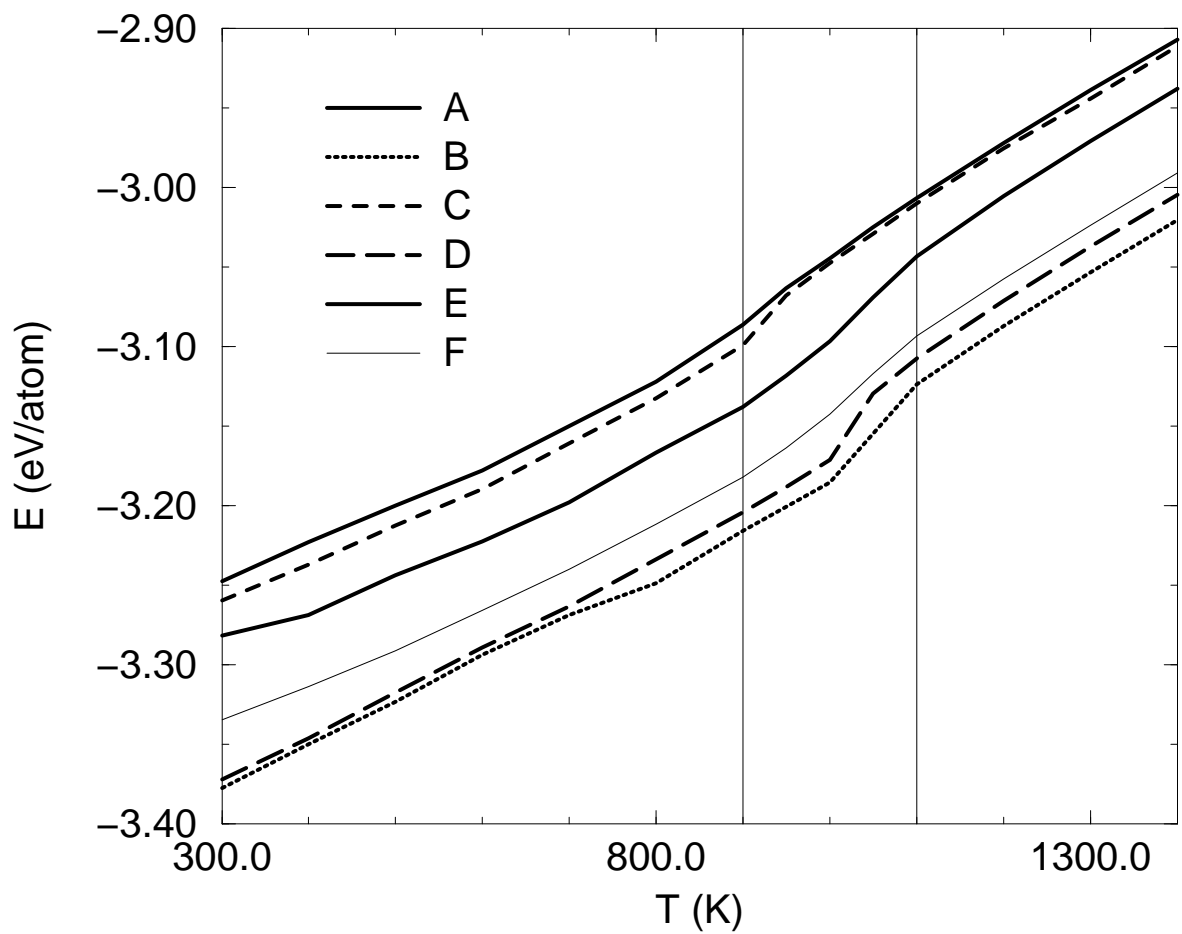
\begin{figure}
\caption{ Caloric functions, i.e., energy as a function of temperature for
all nanowires.
The vertical lines enclose the temperature interval $(900-1100)$ K where
nanowires melt by simultaneous melting of all shells.}
\label{fig5}
\end{figure}


\begin{references}
\bibitem{Landman}  U. Landman, R. N. Barnett, and W. D. Luedtke, Z. Phys. D 
{\bf 40}, 282 (1997).

\bibitem{Ohnishi}  H. Ohnishi, Y. Kondo, and K. Takayanagi, 
Nature {\bf 395}, 780 (1998).

\bibitem{Leiden}  A. I. Yanson, G. Rubio Bollinger, H. E. van den Brom, N.
Agrait, and J. M. van Ruitenbeek, Nature {\bf 395}, 783 (1998).

\bibitem{Pawlow} P. Pawlow, Z. Phys. Chem. {\bf 65}, 545 (1909).

\bibitem{Borel} J. P. Borel, Surf. Sci. {\bf 106}, 1 (1981).

\bibitem{Martin}  T. P. Martin, Phys. Rep. {\bf 273}, 201 (1996).

\bibitem{Bogomolov}  V. N. Bogomolov, E. V. Kolla, and Yu. A. Kumzerov, JETP
Lett. {\bf 41}, 34 (1985).

\bibitem{Yannouleas}  C. Yannouleas and U. Landman, 
J. Phys. Chem. B {\bf 101}, 5780 (1997).

\bibitem{Yanson}  A. I. Yanson, I. K. Yanson, and J. M. van Ruitenbeek,
Nature {\bf 400}, 144 (1999).


\bibitem{Goranka}  G. Bilalbegovi{\' c}, Phys. Rev. B {\bf 58}, 15412
(1998).


\bibitem{Oguz}  O. G{\" u}lseren, F. Ercolessi, and E. Tosatti, Phys.
Rev. Lett. {\bf 80}, 3775 (1998).

\bibitem{Gulseren}  O. G{\" u}lseren, F. Ercolessi, and E. Tosatti,
Phys. Rev. B {\bf 51}, 7377 (1995).

\bibitem{Finbow}  G. M. Finbow, R. M. Lynden-Bell, and I. R. McDonald, Mol.
Phys. {\bf 92}, 705 (1997).

\bibitem{Furio}  F. Ercolessi, M. Parrinello, and E. Tosatti, Philos. Mag. A 
{\bf 58}, 213 (1988).

\bibitem{Portal}  D. Sanchez-Portal, E. Artacho, J. Junquera, P. Ordejon, A.
Garcia, and J. M. Soler, 
Phys. Rev. Lett. {\bf 83}, 3884 (1999). 

\bibitem{DeMaria}  L. De Maria and M. Springborg, preprint, cond-mat/9904181.

\bibitem{Ercolessi}  F. Ercolessi, E. Tosatti, and M. Parrinello, 
Phys. Rev. Lett. {\bf 57}, 719 (1986).

\bibitem{Erio}
G. Bilalbegovi{\'c } and E. Tosatti,
Phys. Rev. {\bf 48}, 11240 (1993).

\bibitem{Tarazona}  P. Tarazona and L. Vicente, Mol. Phys. {\bf 56}, 557
(1985).


\bibitem{Landau}  L. D. Landau and E. M. Lifshitz, 
{\it Statistical Physics}, Pergamon, Oxford (1980).

\bibitem{Scalapino}  Y. Imry and D. J. Scalapino, Phys. Rev. A {\bf 9}, 1672
(1974).

\bibitem{Imry}  Y. Imry, Phys. Rev. B {\bf 21}, 2042 (1980).

\bibitem{Binder} K. Binder, Rep. Prog. Phys. {\bf 50}, 783 (1987). 

\bibitem{www} WebElements,
http://www.shef.ac.uk/$\sim$ chem/webelements/, University of Sheffield,
Shefield (1999).

\end{references}
\end{document}